\documentstyle[aps,multicol,epsf]{revtex}
\begin{document}
\title{
Local non-equilibrium distribution of charge carriers in a phase-coherent
conductor}
\author{Thomas Gramespacher and Markus B\"uttiker}
\address{D\'epartement de Physique Th\'eorique, Universit\'e 
de Gen\`eve, CH-1211, Gen\`eve 4, Switzerland} 
\date{\today}
\maketitle
\begin{abstract}
{\bf abstract} --- We use the scattering matrix approach to derive generalized
Bardeen-like formulae
for the conductances between the contacts of a phase-coherent multiprobe
conductor and a tunneling tip which probes its surface.
These conductances are proportional to
local partial densities of states, called injectivities and
emissivities. The current and the current fluctuations measured at the tip
are related to an {\em effective local non-equilibrium distribution function}.
This distribution function contains the quantum-mechanical phase-coherence of
the charge carriers in the conductor and is given as products of injectivities
and the Fermi distribution functions in the electron reservoirs. The results
are illustrated for measurements on ballistic conductors with barriers and for
diffusive conductors.
\end{abstract}
\vspace{0.5cm}
{\sc Introduction} ---
Over the last 15 years the scanning tunneling microscope (STM)
has developed
into a standard tool to measure the
electronic structure on the surfaces of conductors.
Atomic resolution and even
manipulation of single atoms on a surface has been achieved in many
laboratories\cite{avouris95}.
Most often STMs are used in a two terminal configuration,
the probe being one contact and the STM tip being the other. 
It has been shown that
at zero temperature
the
two-terminal conductance between tip and surface is proportional
to the local density of states (LDOS) on the surface at the coupling
point $x$ of the tip, $\nu(x)$, and given by the Bardeen formula\cite{tersoff},
$G_{tip,surf}=(e^2/h)4\pi^2\nu_{tip}|t|^2\nu(x)$. Here, $\nu_{tip}$ is the
density of states in the tip and $t$ is the coupling strength of the tip
to the surface. 

\begin{figure}
\centerline{\parbox{7cm}{
\epsfxsize7cm
\epsffile{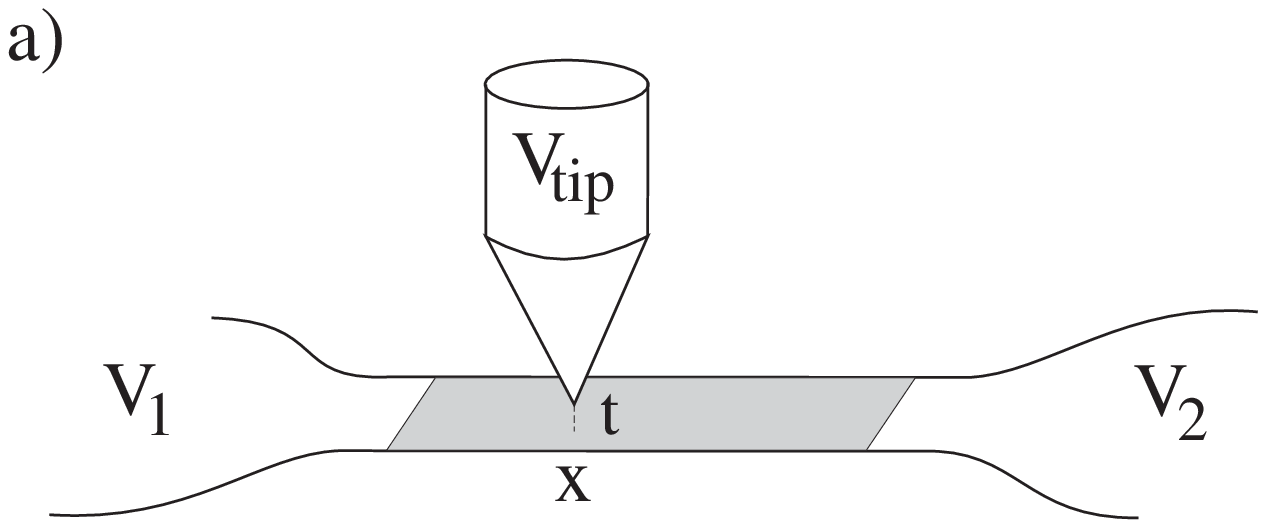}}
\hspace{1.5cm}
\parbox{7cm}{
\epsfxsize7cm
\epsffile{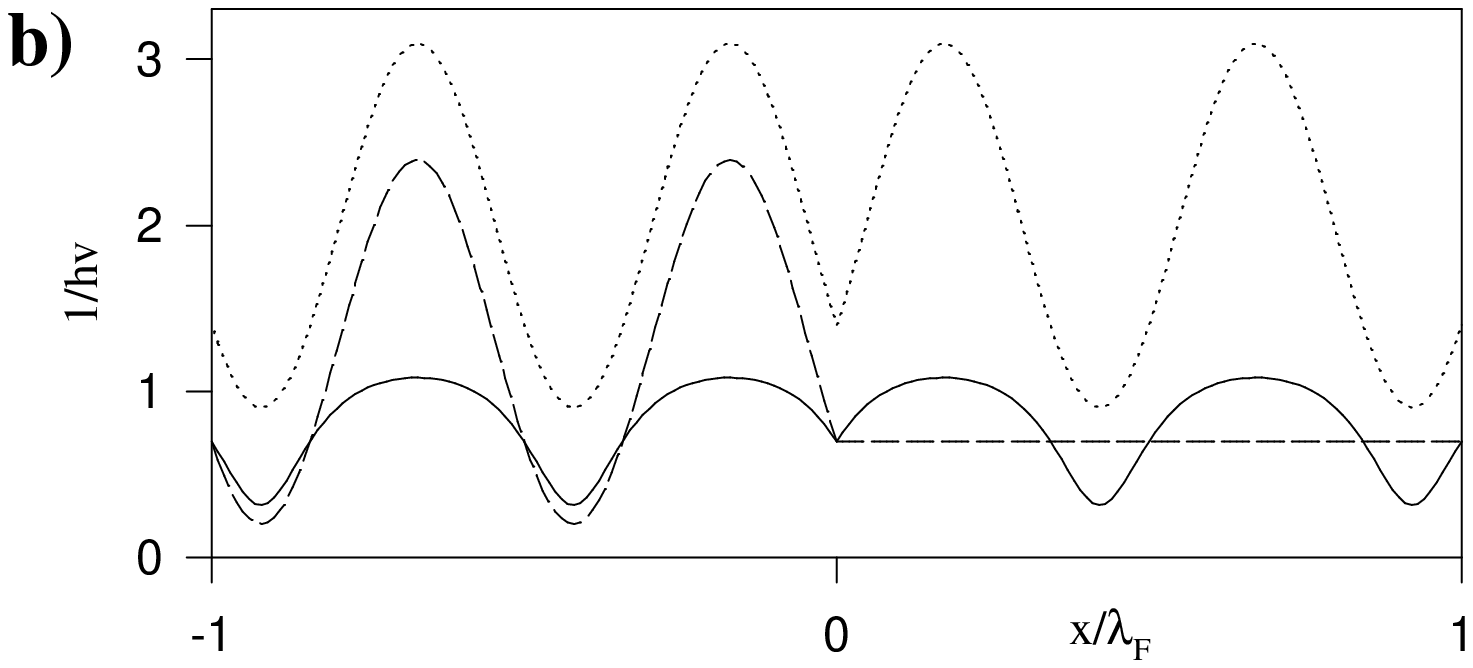}}}
\vspace{0.2cm}
\caption{a) Experimental setup of an STM used to measure on a mesoscopic
multiprobe conductor. The wire is biased with potentials $V_1$ and $V_2$, the
applied potential at the tip is $V_{tip}$. This setup can be used to measure
the average current or the fluctuation spectrum of the current flowing into
the tip. b) Spatial variation of the current fluctuations and L(P)DOS of a
ballistic
wire with a $\delta$ barrier at $x=0$ with transmission probability $T=0.7$.
The injectivity of the left contact, $\nu(x,1)$, (dashed line), and the LDOS
$\nu(x)$, (dotted line) are in units of $1/hv$. The solid line gives the
current fluctuation spectrum $\langle(\Delta I_{tip})^2\rangle$.
}
\label{eintip}
\end{figure}
It is the purpose of this work to suggest and theoretically investigate
experiments which use an STM
to measure
on multiprobe conductors as depicted in Fig.\ 1a).
By applying different
voltages at the contacts of the conductor one can drive a current through the
sample already without the presence of a tunneling contact. Transport through
multiprobe
samples is characterized by a conductance matrix $G_{\alpha\beta}=(e^2/h)
T_{\alpha\beta}$
describing the conductances between each two contacts $\alpha$ and $\beta$
of the sample. In the presence of a tunneling contact the overall
conductance matrix
of the sample plus tip now 
includes the conductances $G_{tip,\alpha}$ from a sample contact to the tip
and the 
conductances $G_{\alpha,tip}$ from the tip to a sample contact. 
In contrast to the typical STM measurement which is characterized 
by one conductance only, we need for the arrangement of Fig. 1a)
to determine two or more conductances. 
One aim of the present work is to investigate the densities
which determine the transmission probabilities $T_{tip,\alpha}$
and $T_{tip,\alpha}$ for transmission from and to the tip. 
Once the conductance matrix is known we can use B\"uttiker's
formula\cite{buttiker86} which expresses the current flowing into the contacts
$\alpha$ of
an arbitrary phase-coherent multiprobe conductor with the help of its 
scattering matrix and the Fermi distribution functions $f_\alpha(E)$
of the electrons 
in the reservoirs\cite{buttiker86},
\begin{equation}
\langle I_\alpha\rangle=\frac{e}{h}\sum_\beta \int dE
T_{\alpha\beta}(E)
\{f_\alpha(E)-f_\beta(E)\}\, .
\label{avcur}
\end{equation}

Contacts weakly coupled to a multiprobe conductor have been investigated 
by Engquist and Anderson\cite{EA}, using scattering matrices to describe 
the splitting of currents between conductor and tip.  
This discussion allowed a derivation of Landauer's resistance
formula\cite{landauer} without appealing to screening arguments 
(local charge neutrality). Only currents 
but not their quantum mechanical amplitudes where considered in 
Ref. \onlinecite{EA}. 
Subsequently, Imry\cite{Imry}  applied the Fermi Golden Rule to 
investigate this problem. 
A fully phase coherent discussion of a weak coupling contact 
(also using scattering matrices) 
was provided by one of us \cite{Butt89}. 
Here we return to this problem. Instead of using scattering matrices which 
describe the coupling of the tip to the sample and a
separate scattering matrix for the 
sample, we start from the overall scattering matrix which includes all
components of the system under investigation, the sample and the tip, 
and use a Green's function technique to arrive at the 
results\cite{gramespacher97}. Such an approach permits readily
to express results either in terms of scattering
matrices or wave functions.

{\sc Average current} ---
We find that the 
transmission probability
from one contact $\alpha$ into the tip is no longer proportional 
to the total LDOS but only to a part of it, a {\it local partial
density of states} (LPDOS) $\nu(x,\alpha)$\cite{buttikerjphys,christen},
called {\it injectivity} of the contact $\alpha$ at the coupling point
$x$. The transmission from the tip into a contact $\alpha$ is proportional to
another LPDOS $\nu(\alpha,x)$, called 
{\it emissivity}\cite{buttikerjphys,Levinson}
of the point $x$ into contact
$\alpha$.
We find\cite{gramespacher97}
\begin{equation}
T_{tip,\alpha}=4\pi^2\nu_{tip}|t|^2\nu(x,\alpha)
\quad\mbox{and}\quad
T_{\alpha,tip}=4\pi^2\nu(\alpha,x)|t|^2\nu_{tip}
\, .
\label{conta}
\end{equation}
These results can be viewed as generalized Bardeen formulae for tunneling
into multiprobe samples.
The injectivity $\nu(x,\alpha)$ can be expressed with the help of the 
scattering states $\psi_{\alpha m}$ which represent an incoming electron
from one of the open channels $m$ of
contact $\alpha$ scattered in all channels of all contacts,
$\nu(x,\alpha)=\sum_{m\in\alpha}(1/hv_{\alpha m})|\psi_{\alpha m}(x)|^2$.
Here, $1/hv_{\alpha m}$ is the density of scattering states in channel $m$ of
contact $\alpha$ and $v_{\alpha m}$ is their Fermi velocity.
The injectivity of a contact $\alpha$ thus gives the density of 
states at the Fermi energy and at the position $x$ for charge carriers which
entered the sample through contact $\alpha$.
Injectivity and emissivity obey the sum rule
$\sum_\alpha \nu(x,\alpha)=\sum_\alpha \nu(\alpha,x)=\nu(x)$
so that Eqs.\ (\ref{conta}) reduce to the well known
Bardeen result in the case where the conductor is only connected to one
single electron reservoir.
Whereas the LDOS $\nu(x)$ is an even function of the magnetic field $B$, 
the injectivity
and emissivity are related to each other by the symmetry relation
$\nu_B(\alpha,x)=\nu_{-B}(x,\alpha)$.
The emissivity can therefore be expressed
with the help of the scattering states of the Hamilton
operator in which the magnetic field 
has been reversed. In addition, this symmetry relation
manifestly shows that the transmission probabilities,
Eqs.\ (\ref{conta}),
satisfy the Onsager-Casimir reciprocity relation\cite{buttiker86},
$T_{\alpha,tip}(B) = T_{tip,\alpha}(-B)$.

Using (\ref{conta}) in Eq. (\ref{avcur})
we can express the current flowing into the
tip in terms of the applied potentials and the injectivities of the sample.
At finite temperature we have
\begin{equation}
\langle I_{tip}\rangle=\frac{e}{h}\int dE T_{ts}(x)\{ f_{tip}(E)-f_{eff}(x)\}
\label{avcurtip}
\end{equation}
with the two-probe tip-to-sample transmission $T_{ts}(x)=4\pi^2
\nu_{tip}|t|^2\nu(x)$ and the {\em effective local distribution
function}
\begin{equation}
f_{eff}(x)=\sum_\alpha\frac{\nu(x,\alpha)}{\nu(x)}f_\alpha(E)\, .
\label{disfunc}
\end{equation}
This expression gives the local non-equilibrium distribution of charge carriers
at
the point $x$ inside the conductor. Its energy dependence comes from the Fermi
distribution functions and a possible energy dependence of the L(P)DOS.
Pothier {\em et al.}\cite{pothier}
measured an over a spatially wide range averaged
non-equilibrium distribution function using a large superconducting tunneling
contact on a diffusive wire.
Eq.\ (\ref{avcurtip}) has the form of the current
in a two probe system, one probe being the tip, where the electron distribution
is described by the Fermi function $f_{tip}(E)$ and an other probe where the
electron distribution is given by the effective distribution function
$f_{eff}(x)$. This effective distribution function does not account for any
energy relaxation of the charge carriers inside the conductor. We assume
that electron-electron and electron-phonon interactions can be neglected
for the system under consideration and therefore the phase of the charge carriers
is conserved. However, the distribution function does
contain via the L(P)DOS the quantum mechanical phase coherence of the electron
wavefunction throughout the system. Our effective
distribution can be used to describe the quasi-particle distribution in
phase-coherent diffusive conductors, if energy relaxation and dephasing can be
neglected. To describe transport and noise in diffusive conductors one can
also use the semiclassical Boltzmann-equation approach, which introduces a
distribution function which does not contain the quantum-mechanical
phase-coherence but where energy relaxation processes can be modeled quite easily.
However, the distribution function of this semiclassical approach can not be
used for conductors where
phase-coherence is essential.

At zero temperature we can replace the Fermi functions in
Eq.\ (\ref{avcurtip})
by step functions and get in linear response to the applied
potentials
\begin{equation}
\langle I_{tip}\rangle=G(x)\{ V_{tip}-V_{eff}(x) \}
\quad\mbox{with}\quad 
V_{eff}(x)=\sum_\alpha\frac{\nu(x,\alpha)}{\nu(x)}V_\alpha
\, ,\label{avcurtiplin}
\end{equation}
and the conductance $G(x)=(e^2/h)T_{ts}(x)$ has to be taken at
the Fermi energy.
A particularly interesting setup is, when the tip is used as a voltage probe,
i.\ e.\ we demand that there is no net current flowing into the tip,
$\langle I_{tip}\rangle =0$.
Similar experiments, also called scanning tunneling potentiometry, have
been reported in \cite{muralt}. 
From Eq. (\ref{avcurtiplin}) we find that at zero temperature
the voltage one has to apply at the tip
to achieve the zero-current condition
is exactly the effective voltage $V_{eff}(x)$ defined in
Eq.\ (\ref{avcurtiplin}). 
The injectivities $\nu(x,\alpha)$ are determined by the equilibrium
electrostatic
potential $U(x)$ in the sample and, therefore, also the measured effective
potential
$V_{eff}(x)$ depends on the electrostatic potential. However, there is
no direct relation between the measured potential
and the actual electrostatic potential in the sample\cite{buttikerjphys}.

In the following we want to evaluate the effective distribution function or
voltage for a metallic diffusive wire and for a wire with a barrier.
First, we consider a diffusive wire of length $L$ to which two electron reservoirs
are attached at the left end (x=0) and at the right end (x=L). 
The injectivity of the left contact (contact 1) and the right contact (contact 2)
is then averaged over many different
disorder configurations
$\nu(x,1)=\nu_0(L-x)/L$ and $\nu(x,2)=\nu_0x/L$
with the constant LDOS $\nu_0$.
This gives for the effective distribution function the known classical result
$f_{eff}(x)=f_1(E)(L-x)/L+f_2(E)x/L$.
It is interesting to contrast this result, with a 
a fully phase-coherent, sample specific result. 
We investigate 
the potential measured on a ballistic one-channel wire with a barrier
at $x=0$ leading to the transmission probability $T$ and reflection probability
$R=1-T$. Such a barrier generates 
strong, Friedel-like oscillations\cite{Butt89,gramespacher97} in the 
neighborhood of the
barrier. To the left of the barrier, the injectivity of the left contact,
contact 1, and right contact, contact 2, are
\begin{equation}
\nu(x,1)=\frac{1}{hv}(1+R+2\sqrt{R}\cos(2k_Fx +\phi)) 
\quad\mbox{and}\quad
\nu(x,2)=\frac{1}{hv}T\, ,\label{injbalbar}
\end{equation}
were $\phi$ is the phase acquired by reflected particles.
At zero temperature this gives the oscillating effective voltage
$V_{eff}(x)=V_1-T(V_1-V_2)/\{2+2\sqrt{R}\cos(2k_Fx+\phi)\}$.
The voltages measured to the left and to the right of a scattering region can
be used to find the four-probe resistance of the scattering region, which
is defined as the measured potential drop divided by the 
current flowing through
the scattering region\cite{buttiker86}.
The measured resistance will in general strongly
depend on the exact positions where the voltage is measured as
the strong oscillations in $V_{eff}(x)$ show.

{\sc Current fluctuations} ---
Until now, we were only interested in the average currents given by
Eq.\ (\ref{avcur}). Now, we want to
go one step further and investigate the fluctuations of the current away from
its average value.
The current fluctuation spectrum can give more information about the
transport properties of a conductor than can be drawn from pure conductance
measurements. Van den Brom and van Ruitenbeek\cite{brom}
used combined conductance
and current fluctuation measurements to characterize the detailed transport
mechanism through few-atom gold contacts. Shot noise measurements have
also been used to identify the fractional charge of the quasiparticles in
the fractional quantum hall regime\cite{glattli}.
The low-frequency current fluctuation, respectively,
correlation spectra are defined as the Fourier transform of the current-current
correlator and given in terms of the scattering matrix
${\bf s}_{\alpha\beta}$ of the system as\cite{noise1}
\begin{equation}
\langle\Delta I_\alpha\Delta I_\beta\rangle=
\int dt\langle \Delta I_\alpha(t+t_0)\Delta I_\beta(t_0)
\rangle
=\frac{2e^2}{h}\sum_{\delta\gamma}\int dE {\rm Tr}[
A_{\delta\gamma}(\alpha)A_{\gamma\delta}(\beta)]f_\delta(1-f_\gamma)\, .
\label{gennoise}
\end{equation}
with $\Delta I_\alpha(t)=I_\alpha(t)-\langle I_\alpha(t)\rangle$
and the current matrix $A_{\beta\gamma}(\alpha)=\delta_{\alpha\beta}
\delta_{\alpha\gamma}-{\bf s}_{\alpha\beta}^\dagger{\bf s}_{\alpha\gamma}$.
If the two indices $\alpha$ and $\beta$ are equal, then this equation gives
the current-fluctuation spectrum at the contact $\alpha$. Otherwise, the 
equation gives the correlation spectrum of the currents at contacts $\alpha$
and $\beta$.
Current fluctuations at the contacts of a mesoscopic conductor have been
measured by many groups\cite{lietc}. Birk {\em et al.}\cite{birk}
used an STM to measure the shot-noise on a tunneling contact. Correlations
of currents at two different contacts of a sample
have been measured only recently by Henny {\em et al.}\cite{henny}
and Oliver {\em et al.}\cite{oliver}.

We calculate the current-fluctuation spectrum at
the tunneling contact for the setup shown in Fig.\ 1a).
As in our discussion of the sample to tip conductances, Eq. (\ref{conta}),
we start with the scattering matrix of the entire system (wire and tip) and
develop the scattering matrix elements describing scattering from and to the tip
to the lowest order in the coupling energy $|t|$ of the tip. We assume that the
potential at the tip is adjusted such that the average current at the tip
vanishes. Then, using $G(x)=(e^2/h)4\pi^2\nu_{tip}|t|^2\nu(x)$,
we find for the fluctuations of the current at the tip 
\begin{equation}
\langle (\Delta I_{tip})^2\rangle = 4\int
dEG(x)\{ 1-f_{tip}(E)\} f_{eff}(x)
\, .
\label{tipfluc}
\end{equation}
We see, that the fluctuations are also determined by the effective distribution
function, Eq.\ (\ref{disfunc}).

At elevated temperatures the fluctuations are due to thermal noise in addition
with an excess noise, called shot-noise, which is due to the discreteness of
the charge carriers.
In Eq.\ (\ref{tipfluc}) the integral over energy is from the bottom of the
conduction band to infinity. At a temperature $T$ and applied potential differences
$\Delta V$, the relevant contribution to the current fluctuations comes from the
integration over an energy intervall of $\Delta E\approx\mbox{max}(e\Delta V,
kT)$ around the Fermi energy. If the L(P)DOS are (nearly) independent of
energy in this energy range, we can evaluate the integral over products
of Fermi functions and get
\begin{equation}
\langle (\Delta I_{tip})^2\rangle=2eG(x)\Delta V\frac{\nu(x,1)}{\nu(x)}
\frac{\nu(x,2)}{\nu(x)}\left\{ \coth\left(\frac{\nu(x,1)}{\nu(x)}\frac{e\Delta
V}{2kT}\right)+\coth\left(\frac{\nu(x,2)}{\nu(x)}\frac{e\Delta V}{2kT}
\right)\right\}
\label{finitem}
\end{equation}
with $\Delta V=V_1-V_2$ the applied potential at the wire. In the limit of
high temperature, $kT\gg e\Delta V$, this equation leads to the Johnson-Nyquist
formula for thermal noise, $\langle(\Delta I_{tip})^2\rangle=4eG(x)kT$.
At zero-temperature, we are dealing with pure shot-noise.
If
we restrict ourselves to small potential differences so that linear response
theory is justified, the fluctuation spectrum is completely determined by the
properties of the system (its LPDOS) at the Fermi-energy\cite{gramespacher98},
\begin{equation}
\langle (\Delta I_{tip})^2 \rangle=4eG(x)\Delta V\frac{\nu(x,1)}{\nu(x)}\left(
1-\frac{\nu(x,1)}{\nu(x)}\right)\, . \label{linnoise}
\end{equation}
Using the injectivities, Eq.\ (\ref{injbalbar}), the
measurement of the current fluctuation spectrum, Eq.\ (\ref{linnoise}), is
illustrated in Fig.~1b) for the case of a measurement on a
ballistic one channel conductor with a $\delta$-barrier at $x=0$ leading to
the transmission probability $T=0.7$.

In conclusion,
we presented generalized Bardeen formulae which for the 
transmission from the sample contacts to the tunneling tip 
are related to a local partial density of states, the
injectivities $\nu(x,\alpha)$, and for the transmission
from the tip to the sample contacts are related to the 
emissivities.
We have shown that measurements of the average current and
the fluctuations of the current at a tunneling tip which probes a
phase-coherent multiprobe sample are determined by
an effective local non-equilibrium distribution function. This distribution
function is given
as products of injectivities
and the equilibrium Fermi distribution functions in the electron
reservoirs.
We illustrated the results for
the case of ballistic conductors with barriers and metallic diffusive wires.

This work was supported by the Swiss National Science Foundation.

\end{document}